\documentclass[preprint,showpacs,preprintnumbers,amsmath,amssymb]{revtex4}
\usepackage{booktabs}
\usepackage{mathrsfs}
\usepackage{epsfig}
\usepackage{graphicx}
\usepackage{dcolumn}
\usepackage{bm}
\usepackage{amsmath}
\usepackage{slashed}
\usepackage{multirow}

\let\jnfont=\rm
\def\NPB#1,{{\jnfont Nucl.\ Phys.\ B }{\bf #1},}
\def\PLB#1,{{\jnfont Phys.\ Lett.\ B }{\bf #1},}
\def\EPJC#1,{{\jnfont Eur.\ Phys.\ Jour.\ C }{\bf #1},}
\def\PRD#1,{{\jnfont Phys.\ Rev.\ D }{\bf #1},}
\def\PRL#1,{{\jnfont Phys.\ Rev.\ Lett.\ }{\bf #1},}
\def\MPLA#1,{{\jnfont Mod.\ Phys.\ Lett.\ A }{\bf #1},}
\def\JPG#1,{{\jnfont J.\ Phys.\ G}{\bf #1},}
\def\CTP#1,{{\jnfont Commun.\ Theor.\ Phys.\ }{\bf #1},}
\def\ZPC#1,{{\jnfont Z.\ Phys.\ C }{\bf #1},}
\def\JHEP#1,{{\jnfont JHEP \ }{\bf #1},}
\def\Rv{\not{\hbox{\kern-1pt $R$}}}
\def\p{\not{\hbox{\kern-3pt $p$}}}

\begin{document}

\title{Single top and Higgs associated production in the minimal $B-L$ model at the LHC }

\author{Bingfang Yang$^{1}$}\email{yangbingfang@htu.edu.cn}
\author{Zhiyong Liu$^{1}$}\email{021168@htu.cn}
\author{Jinzhong Han$^{2}$}\email{hanjinzhong@zknu.edu.cn}
\author{Guang Yang$^{3}$}\email{yxd5460@163.com}
\affiliation{ $^1$ College of Physics $\&$ Electronic Engineering,
Henan Normal University, Xinxiang 453007, China\\$^2$School of
Physics and Electromechnical Engineering, Zhoukou Normal University,
Henan, 466001, China\\ $^3$Basic Teaching Department, Jiaozuo
University, Jiaozuo 454000, China
   \vspace*{1.5cm} }%

\date{\today}

\begin{abstract}

In this paper, we study the single top production in association
with a Higgs boson in the $U(1)_{B-L}$ extension of the Standard
Model at the LHC. We calculate the production cross sections of the
processes $pp\rightarrow thX(h=H_{1},H_{2};X=j,b,W)$ in this model.
Then we further study the observability of the process
$pp\rightarrow tH_{2}j$ through $pp\rightarrow t(\rightarrow
q\bar{q'} b)H_{2}(\rightarrow 4\ell)j$. We find that the systematic
significance can be improved obviously, but it is still challenging
for the 14 TeV LHC with high-luminosity to detect this signal.

\end{abstract}
\pacs{14.65.Ha,14.80.Ly,11.30.Hv} \maketitle

\section{INTRODUCTION}
In July 2012, a Higgs-like resonance with mass $m_{h}\sim$ 125 GeV
has been caught by the ATLAS and CMS experiments at the Large Hadron
Collider (LHC)\cite{LHC-higgs1}. So far, all the measurements of the
discovered new particle\cite{LHC-higgs2} are well compatible with
the scalar boson predicted by the Standard
Model(SM)\cite{LHC-higgs3}.

It is well known that the SM cannot be the final theory of nature.
Theoretically, successful explanation of some problems, such as the
hierarchy problem, requires new physics beyond the SM near the TeV
scale. Experimentally, the solid evidence for neutrino oscillation
is one of the firm hints for new physics. The minimal extension of
the SM is that the SM gauge groups are augmented by a $U(1)_{B-L}$
symmetry, where $B$ and $L$ represents the baryon number and lepton
number respectively. The $B-L$ gauge symmetry can explain the
presence of three right-handed neutrinos and provide a natural
framework for the seesaw mechanism\cite{seesaw}. In addiction, it's
worth noting that $B-L$ symmetry breaking takes place at the TeV
scale, hence giving rise to new and interesting TeV scale
phenomenology.

Concerning the probe of new physics through the Higgs boson, the
Yukawa couplings play an important role in probing the new physics.
The top quark is the heaviest particle discovered and owns the
strongest Yukawa coupling. The top quark Yukawa coupling is
speculated to be sensitive to the electroweak symmetry breaking
(EWSB) mechanism and new physics. The $t\bar{t}h$ production process
is a golden channel for directly probing the top Yukawa coupling,
however, this process cannot provide the information on the relative
sign between the coupling of the Higgs to fermions and to vector
bosons. As a beneficial supplement, the $thj$ production process can
bring a unique possibility\cite{thj-theory} and many relevant works
have been carried out\cite{thj-work}.

The $U(1)_{B-L}$ model predicts heavy neutrinos, a TeV scale extra
neutral gauge boson and an additional heavy neutral Higgs, which
makes the model phenomenologically rich. The heavy Higgs state mixes
with the SM Higgs boson so that some Higgs couplings are modified
and this effect can also influence the process of single top and
Higgs associated production. Besides, the process of single top and
heavy Higgs associated production deserves attention, which is
equally important for understanding the EWSB and probing new
physics. By performing the detailed analysis on this process may
provide a good opportunity to probe the $U(1)_{B-L}$ model signal.

The paper is structured as follows. In Sec.II we review the
$U(1)_{B-L}$ model related to our work. In Sec.III we first
calculate the production cross sections of the single top and
$h(=H_{1},H_{2})$ associated production at the LHC, then explore the
observability of $t$-channel process $pp\rightarrow tH_{2}j$ through
$pp\rightarrow t(\rightarrow q\bar{q'} b)H_{2}(\rightarrow 4\ell)j$
by performing a parton-level simulation. Finally, we make a summary
in Sec.IV.

\section{A brief review of the $U(1)_{B-L}$ model}

The minimal $B-L$ extension of the SM \cite{BL-model} is based on
the gauge group $SU(3)_{c}\times SU(2)_{L}\times U(1)_{Y}\times
U(1)_{B-L}$ with the classical conformal symmetry. Under this gauge
symmetry, the invariance of the lagrangian implies the existence of
a new gauge boson. In order to make the model free from all the
gauge and gravitational anomalies, three generations of right-handed
neutrinos are necessarily introduced.

In this model, the most general gauge-invariant and renormalisable
scalar Lagrangian can be expressed as
\begin{eqnarray}
\mathcal{L}_s=(D^{\mu}H)^{\dagger}D_{\mu}H+(D^{\mu}\chi)^{\dagger}(D_{\mu}\chi)-V(\chi,
H),
\end{eqnarray}
with the scalar potential given by
\begin{eqnarray}
V(\chi, H)=M_H^2 H^{\dagger}H+ m^2_{\chi} |\chi|^2 + \lambda_1
(H^{\dagger} H)^2+ \lambda_2 |\chi|^4 + \lambda_3 (H^{\dagger}
H)|\chi|^2. \label{scpot}
\end{eqnarray}

From the mass terms in the scalar potential, the mass matrix between
the two Higgs bosons in the basis $(H,\chi)$ can be given by
\begin{equation}
M^2 (H,\chi) =2\left(%
\begin{array}{cc}
  \lambda_1 v^2 & \frac{\lambda_3}{2} v v' \\
 \frac{\lambda_3}{2} v v' & \lambda_2 v'^2 \\
\end{array}%
\right). %
\end{equation}
The mass eigenstates are related via the mixing matrix
\begin{equation}
\left(\begin{array}{c}
H_1 \\
H_2
\end{array}\right)= \left(\begin{array}{cc}
\cos \alpha & \sin \alpha  \\
-\sin \alpha & \cos \alpha
\end{array}\right)
\left(\begin{array}{c}
H \\
\chi
\end{array}\right),
\label{mixingh}
\end{equation}
where the mixing angle $\alpha$ ($-\frac{\pi}{2} < \alpha <
\frac{\pi}{2}$) satisfies
\begin{equation}
\tan 2\alpha= \frac{ \lambda_3 v' v}{(\lambda_2 v'^2 -\lambda_1
v^2)}. \label{theta}
\end{equation}
The masses of the physical Higgs bosons $H_1$ and $H_2$ are given by
\begin{equation}
m^2_{H_{1},H_{2}} = \lambda_1 v^2 + \lambda_2 v'^2 \mp
\sqrt{(\lambda_1 v^2 - \lambda_2 v'^2)^2 +(\lambda_3 v
v')^{2}},\end{equation} where $H_{1}$ and $H_{2}$ are light SM-like
and heavy Higgs bosons, respectively.

To complete the discussion on the Lagrangian, we write down the
Yukawa term, which in addition to the SM terms has interactions
involving the right-handed neutrinos $N_R$,
\begin{align}
 \mathcal{L}_Y=&-y^d_{ij} \overline{(Q_{L})_i} (d_{R})_j H - y^u_{ij} \overline{(Q_{L})_i} (u_{R})_j \widetilde{H}
 - y^e_{ij} \overline{(L_{L})_i} (e_{R})_j H \nonumber \\ &- y^{\nu}_{ij} \overline{(L_{L})_i} (N_{R})_j \widetilde{H}
 - y^M_{ij} \overline{(N_{R})_i^c} (N_{R})_j \chi + h.c.,
\end{align}
where $\widetilde{H}=i\sigma^2 H^*$ and $i,j$ runs from 1$\sim $3.
The vacuum expectation value (VEV) of the $\chi$ field breaks the
$B-L$ symmetry and generates the Majorana masses for the right
handed neutrinos and the Dirac masses for the light neutrinos.

In terms of the mixing angle $\alpha$, the couplings of $H_1$ and
$H_2$ with the fermions and gauge bosons can be expressed as follows
\begin{eqnarray}
&&H_1 f \bar{f} : - \frac{e M_f \cos \alpha}{2 M_Ws_W}, ~~~~~ \hspace*{0.5cm} H_2 f \bar{f} : - \frac{e M_f \sin \alpha}{2 M_Ws_W},  \nonumber \\
&&H_1 W^{+} W^{-} : \frac{e M_W  \cos \alpha}{s_W},  \hspace*{0.5cm} H_2 W^{+} W^{-} : \frac{e M_W  \sin \alpha}{s_W}, \nonumber \\
&&H_1 Z Z : \frac{e M_W  \cos \alpha}{c^2_W s_W}, ~~~~~
\hspace*{0.5cm} H_2 Z Z :\frac{e M_W  \sin \alpha}{c^2_W s_W}.
\label{int}
\end{eqnarray}
where $f$ denotes the SM fermions, $s_W=\sin\theta_W$ with
$\theta_{W}$ is the usual Weinberg angle.

\section{Numerical results and discussions}
\begin{figure}[htbp]
\scalebox{0.55}{\epsfig{file=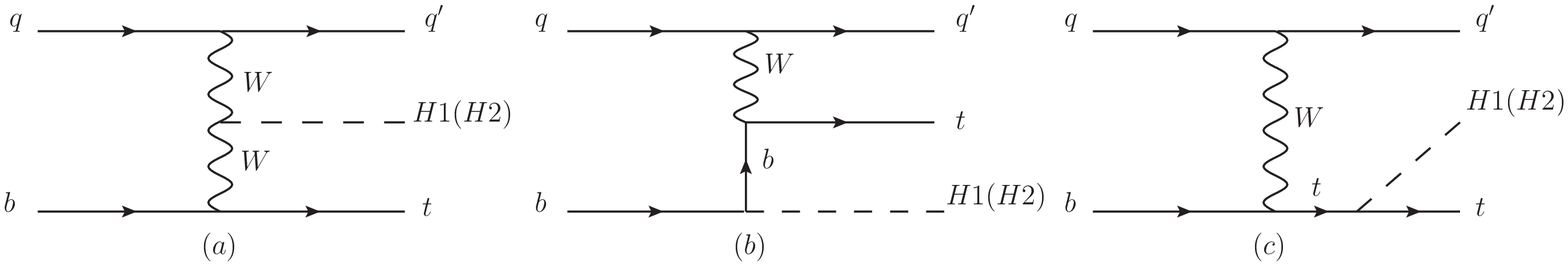}}\vspace{-0.5cm}\caption{
Lowest-order feynman diagrams for $pp \to tH_{1}(H_{2})j$ in the
$U(1)_{B-L}$ model.}\label{thjBL}
\end{figure}

\begin{figure}[htbp]
\scalebox{0.4}{\epsfig{file=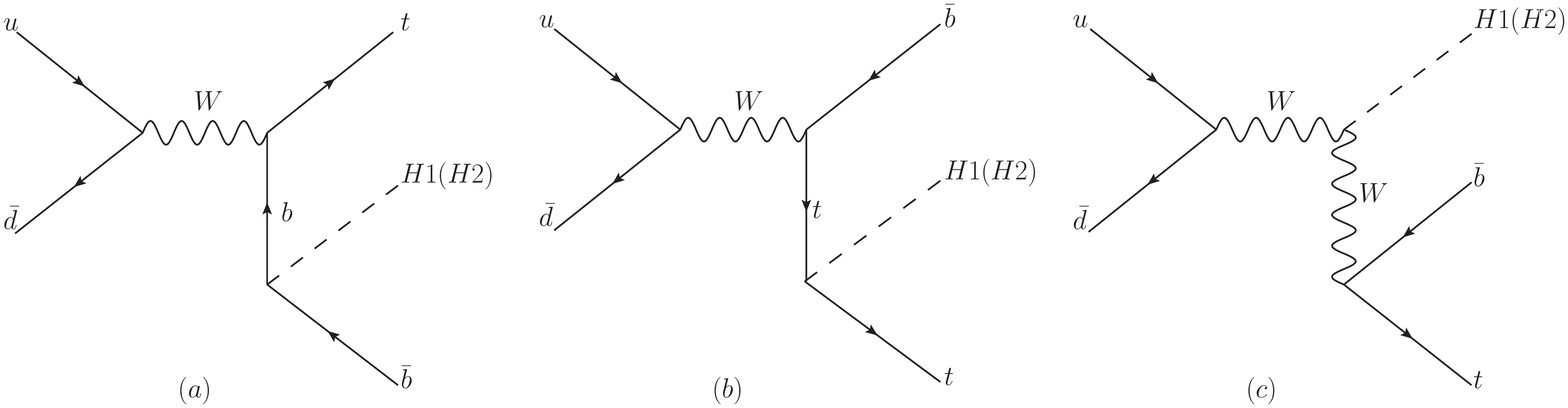}}\vspace{-0.5cm}\caption{
Lowest-order feynman diagrams for $pp \to tH_{1}(H_{2})\bar{b}$ in
the $U(1)_{B-L}$ model.}\label{thbBL}
\end{figure}

\begin{figure}[htbp]
\scalebox{0.4}{\epsfig{file=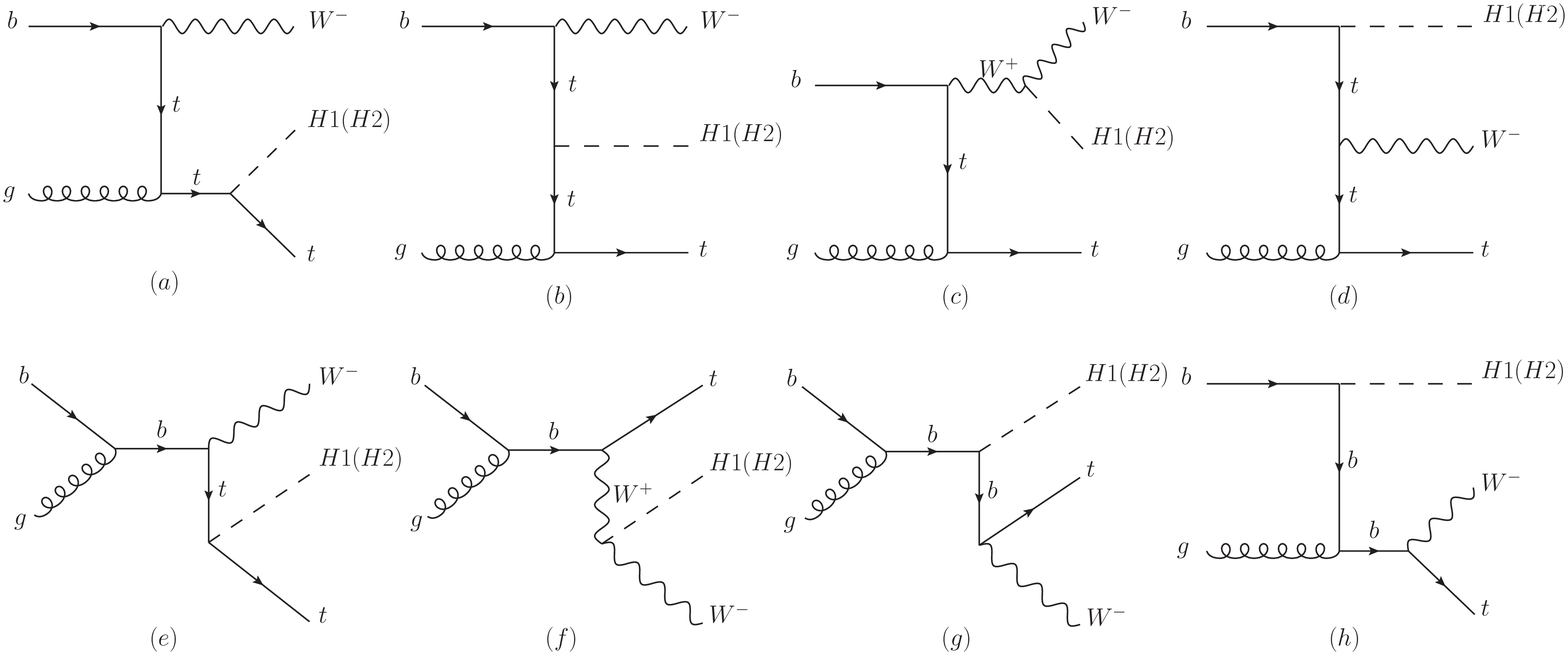}}\vspace{-0.5cm}\caption{
Lowest-order feynman diagrams for $pp \to tH_{1}(H_{2})W^{-}$ in the
$U(1)_{B-L}$ model.}\label{thwBL}
\end{figure}

For the single top and Higgs associated production, the three
processes of interest are characterized by the virtuality of the $W$
boson in the process\cite{thj-modes}: (i) $t$-channel, where the $W$
is spacelike; (ii) $s$-channel, where the $W$ is timelike; (iii)
$W$-associated production channel, where there is emission of a real
$W$ boson. In the $U(1)_{B-L}$ model, the lowest-order Feynman
diagrams of the $t$-channel process $pp \to tH_{1}(H_{2})j(j\neq b)$
are shown in Fig.\ref{thjBL}, the $s$-channel process $pp \to
tH_{1}(H_{2})\bar{b}$ are shown in Fig.\ref{thbBL} and the
$W$-associated production channel process $pp \to
tH_{1}(H_{2})W^{-}$ are shown in Fig.\ref{thwBL}. We can see that
the Feynman diagrams for these processes are the same as the
corresponding SM processes. Moreover, the conjugate processes where
$t$ is replaced by $\bar{t}$ have been included in our calculations.

We compute the cross sections by using $\textsf{CalcHEP
3.6.25}$\cite{calchep} with the parton distribution function CTEQ6L
\cite{cteq}, and set the renormalization scale $\mu_R$ and
factorization scale $\mu_F$ to be
$\mu_R=\mu_F=(m_{t}+m_{h}+m_{X})/2, (h=H_{1},H_{2}; X=j,b,W)$. The
SM input parameters are taken as follows \cite{pdg}:
\begin{align}
m_t = 173.2{\rm ~GeV},\quad &m_{Z} =91.1876 {\rm ~GeV}, \quad
\alpha(m_Z) = 1/128, \\ \nonumber \sin^{2}\theta_W = 0.231,\quad
&m_{H_{1}} =125 {\rm ~GeV}, \quad \alpha_{s}(m_Z)=0.1185.
\end{align}

In our calculations, the relevant $U(1)_{B-L}$ model parameters are
the mixing parameter $\alpha$ and the heavy Higgs mass $m_{H_{2}}$.
Considering the constraints in Refs.\cite{constraint,constraint2},
we choose the parameter space as follows: $0.01<\sin\alpha<0.4$,
$250\rm GeV$$ \leq m_{H_{2}}\leq$$ 1000\rm GeV$.
\subsection{Single top and $H_{1}$ associated production}

\begin{figure}[htbp]
\scalebox{0.35}{\epsfig{file=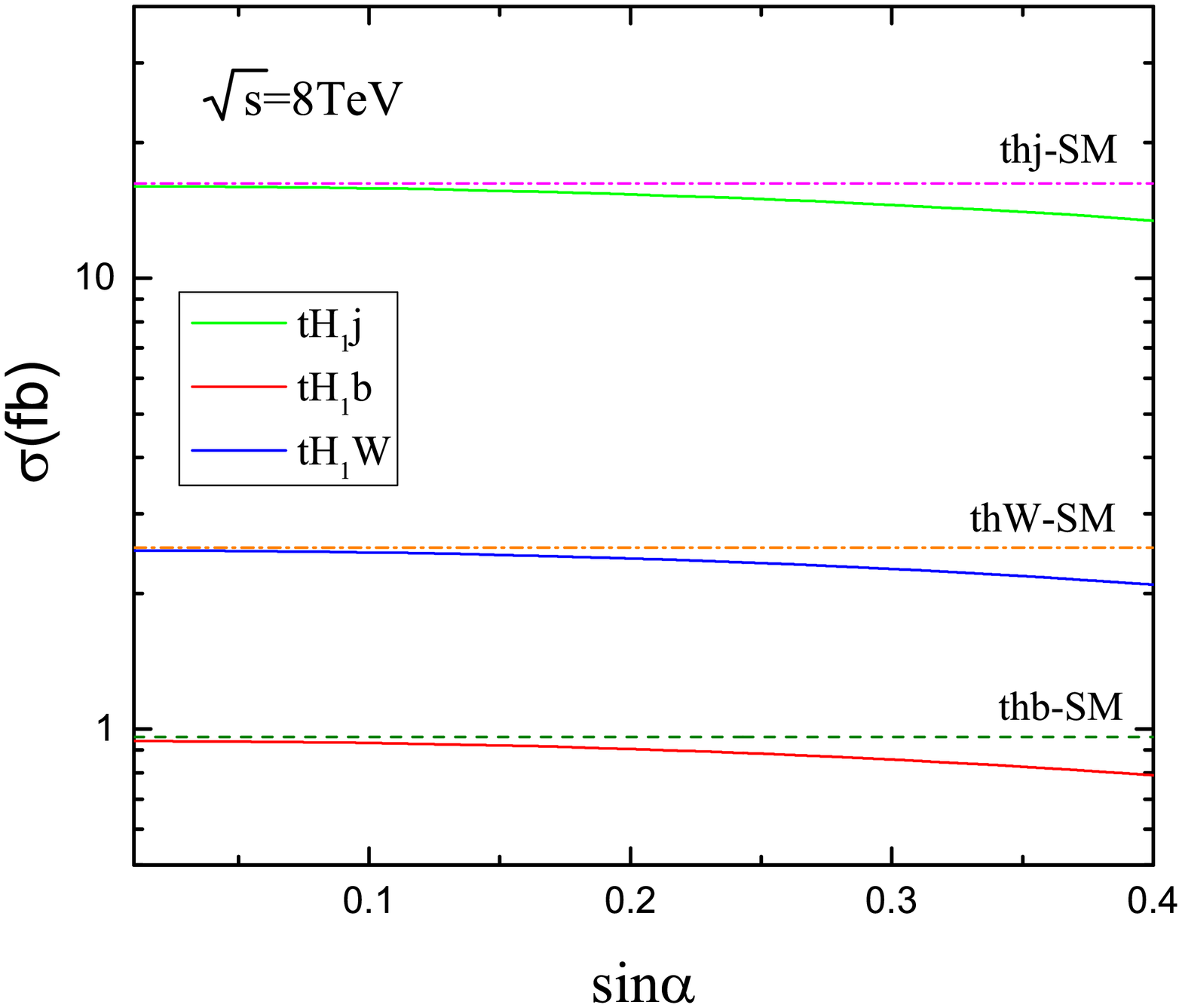}}\vspace{-0.5cm}\hspace{-1.5cm}
\scalebox{0.35}{\epsfig{file=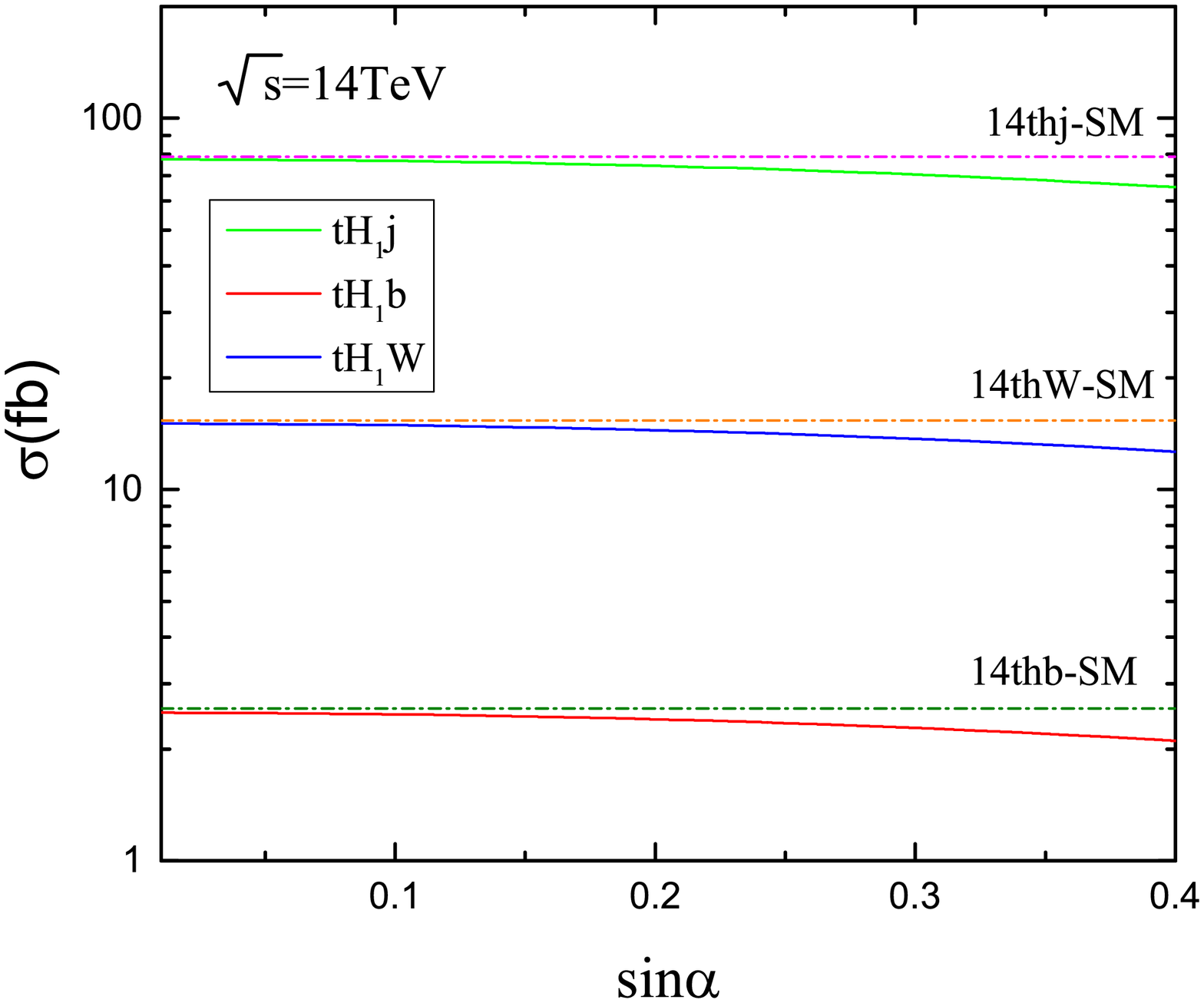}}
\vspace{-0.5cm}\caption{The production cross sections
$\sigma_{tH_{1}j}, \sigma_{tH_{1}b}, \sigma_{tH_{1}W}$ as a function
of $\rm sin\alpha $ at 8,14 TeV LHC in the $U(1)_{B-L}$
model.}\label{th1}
\end{figure}

In Fig.\ref{th1}, we show the production cross sections of the
processes $pp \to tH_{1}j$, $pp \to tH_{1}\bar{b}$ and $pp \to
tH_{1}W^{-}$ as a function of $\rm sin\alpha $ at the 8 and 14 TeV
LHC in the $U(1)_{B-L}$ model, respectively. For clarity, we marked
the corresponding SM process cross sections on the figures. We can
see that the cross sections in the $U(1)_{B-L}$ model decrease with
increasing $\sin\alpha$. Besides, the behavior of these three
processes are similar for the 8 TeV and 14 TeV. This is easy to
understand because there are the same change factor $\cos\alpha$ in
the light Higgs $H_{1}$ couplings in Eq.(\ref{int}) so that the
production cross sections are suppressed by $\cos^{2}\alpha$, i.e.,
$\sigma_{B-L}=\sigma_{SM}\cos^{2}\alpha$. When $\sin\alpha
\rightarrow 0$, the mixing between the light Higgs $H_{1}$ and the
heavy Higgs $H_{2}$ will decouple so that the cross sections go back
to the SM values.

\subsection{Single top and $H_{2}$ associated production}
\begin{figure}[htbp]
\scalebox{0.36}{\epsfig{file=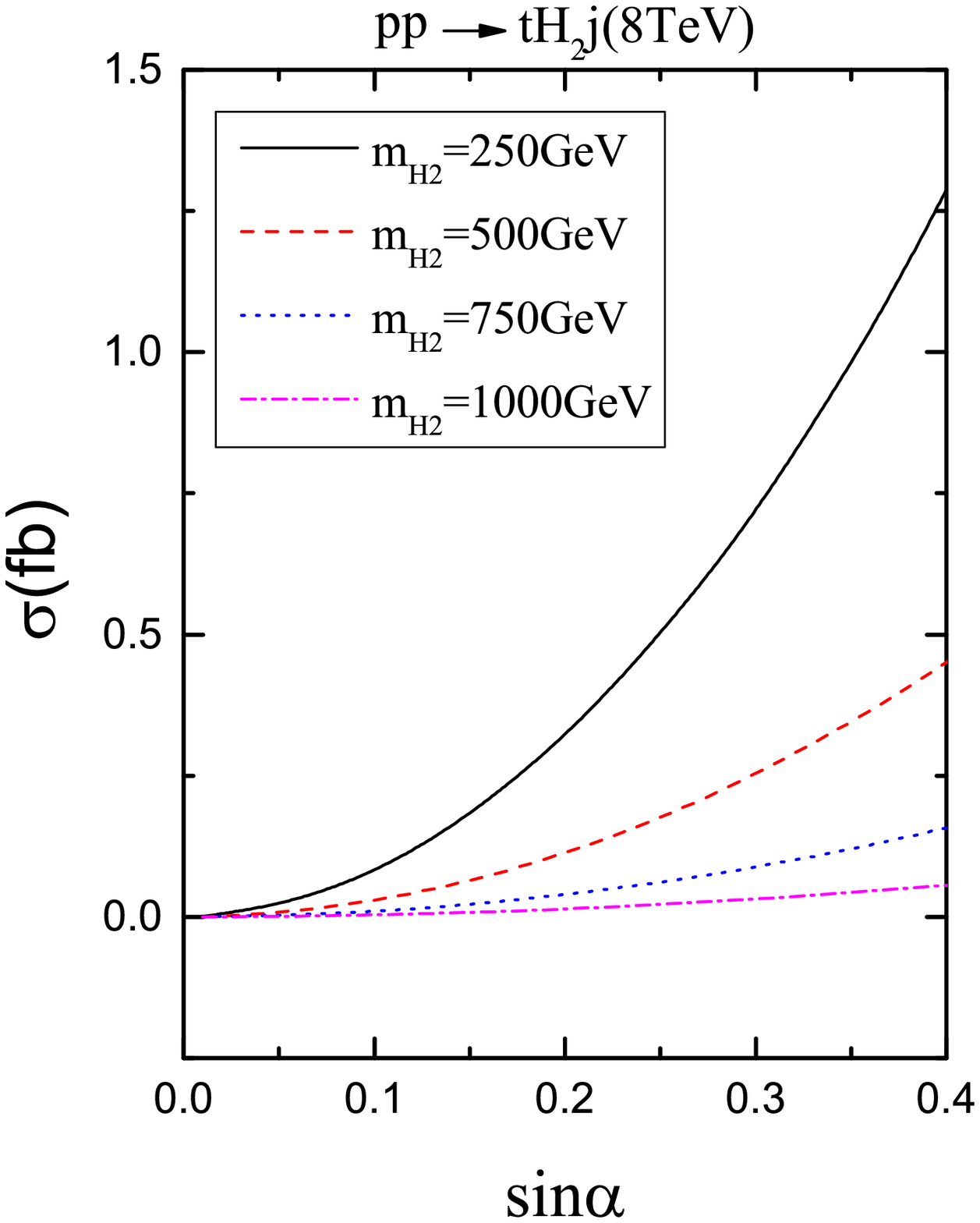}}\vspace{-0.5cm}\hspace{-0.9cm}
\scalebox{0.36}{\epsfig{file=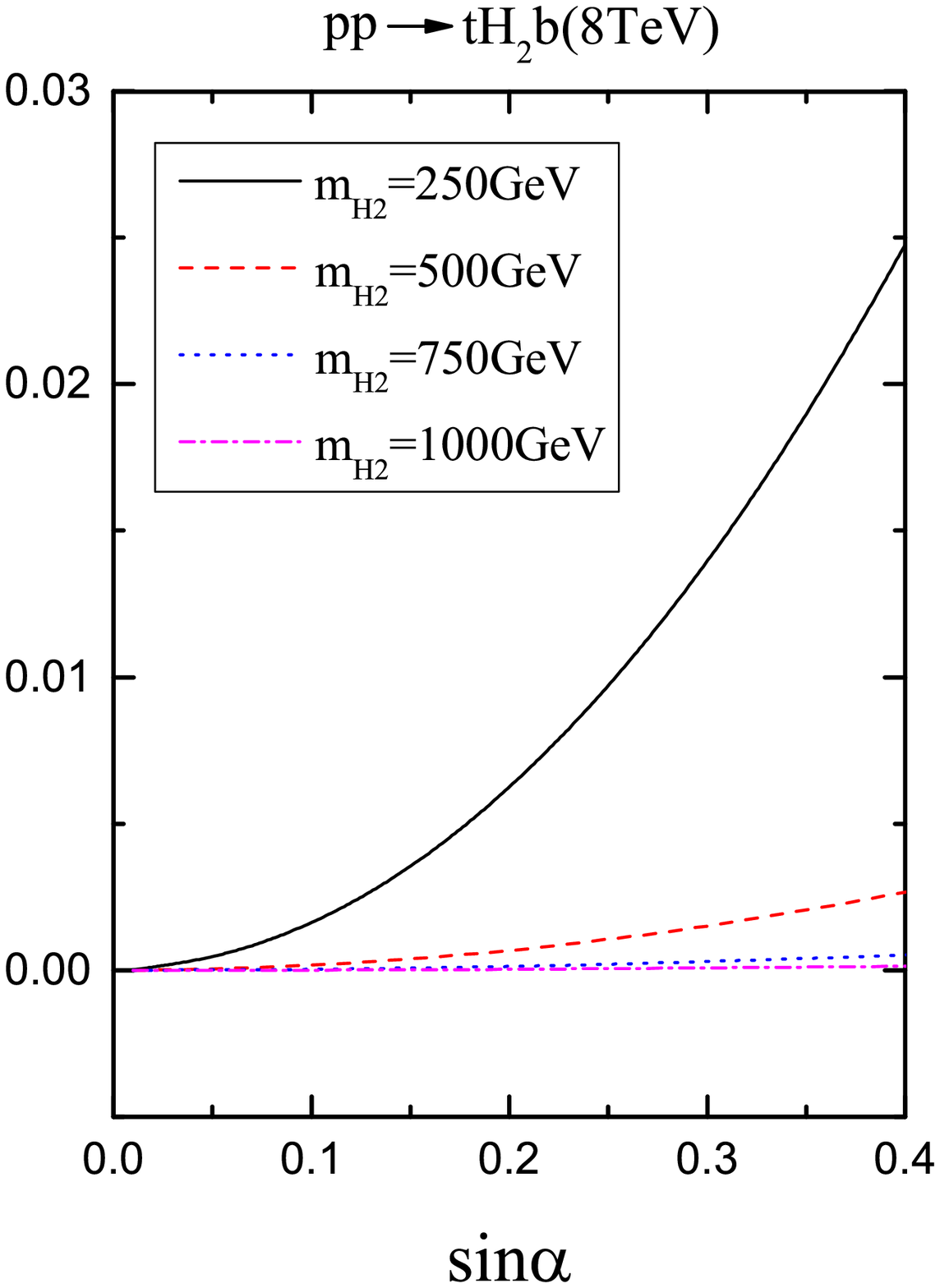}}\vspace{-0cm}\hspace{-0.9cm}
\scalebox{0.36}{\epsfig{file=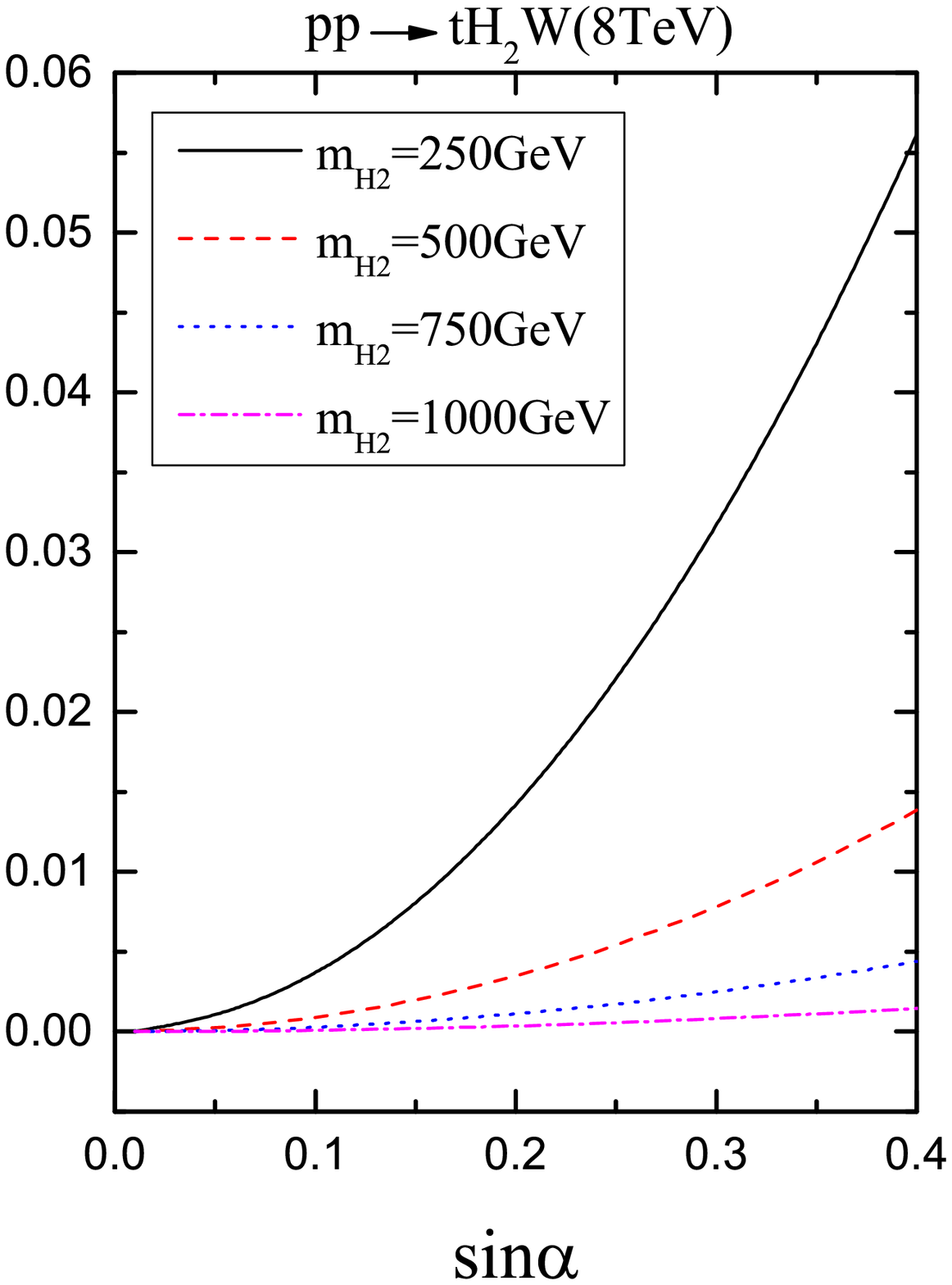}}
\vspace{-0cm}\caption{The production cross sections
$\sigma_{tH_{2}j}, \sigma_{tH_{2}b}, \sigma_{tH_{2}W}$ as a function
of $\rm sin\alpha $ at 8 TeV LHC in the $U(1)_{B-L}$
model.}\label{th2cross8}
\end{figure}
\begin{figure}[htbp]
\scalebox{0.36}{\epsfig{file=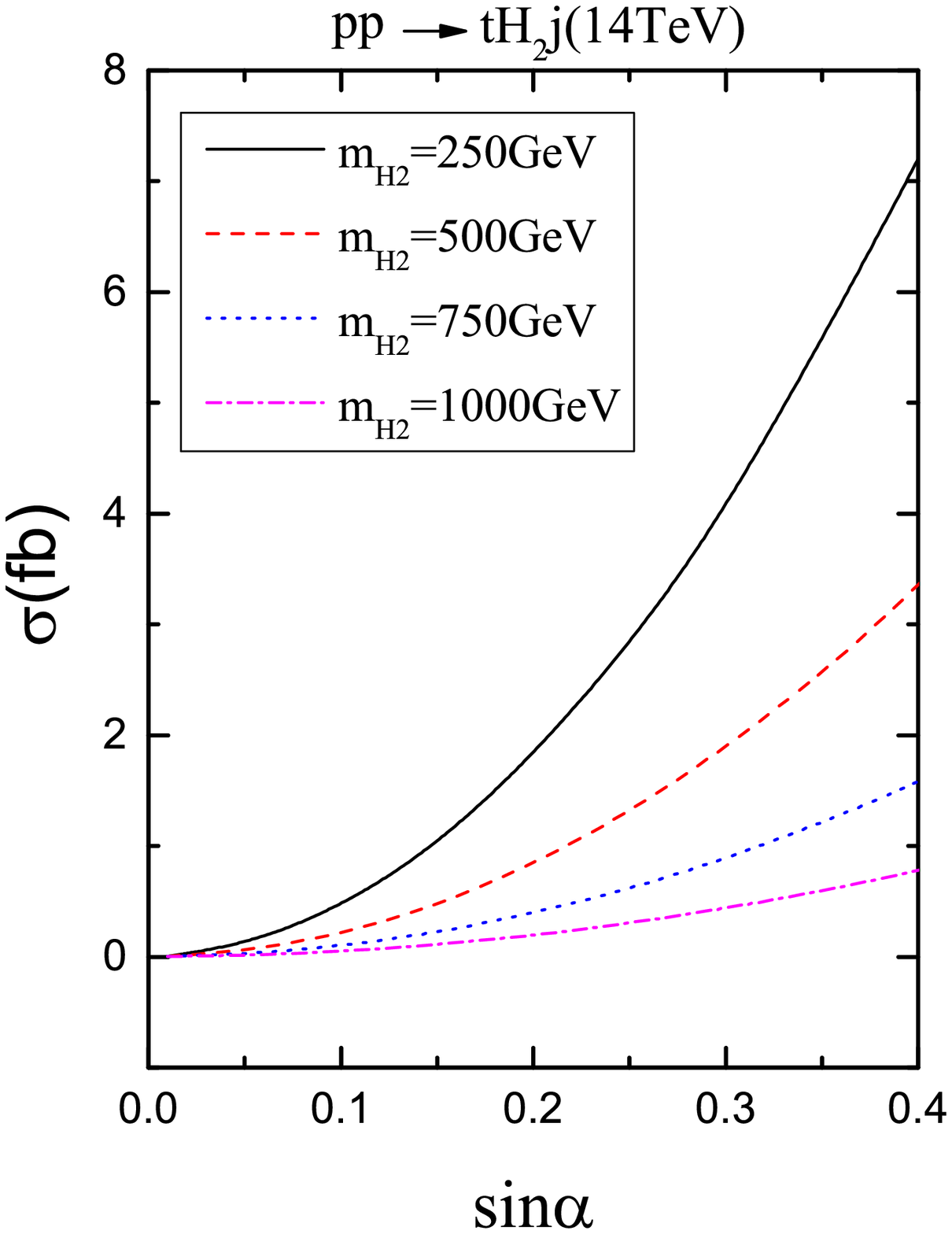}}\vspace{-0.5cm}\hspace{-0.9cm}
\scalebox{0.36}{\epsfig{file=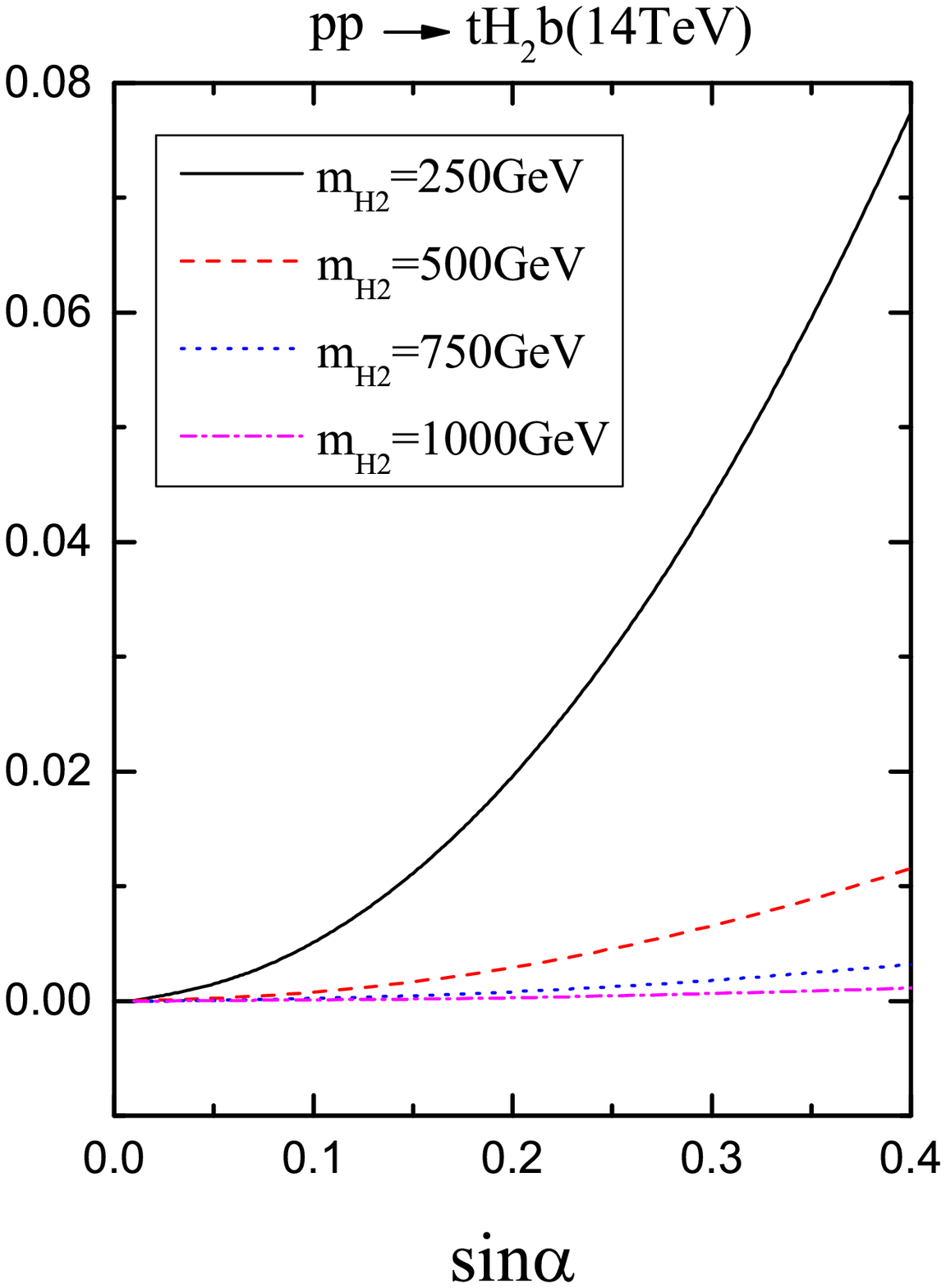}}\vspace{-0cm}\hspace{-0.9cm}
\scalebox{0.36}{\epsfig{file=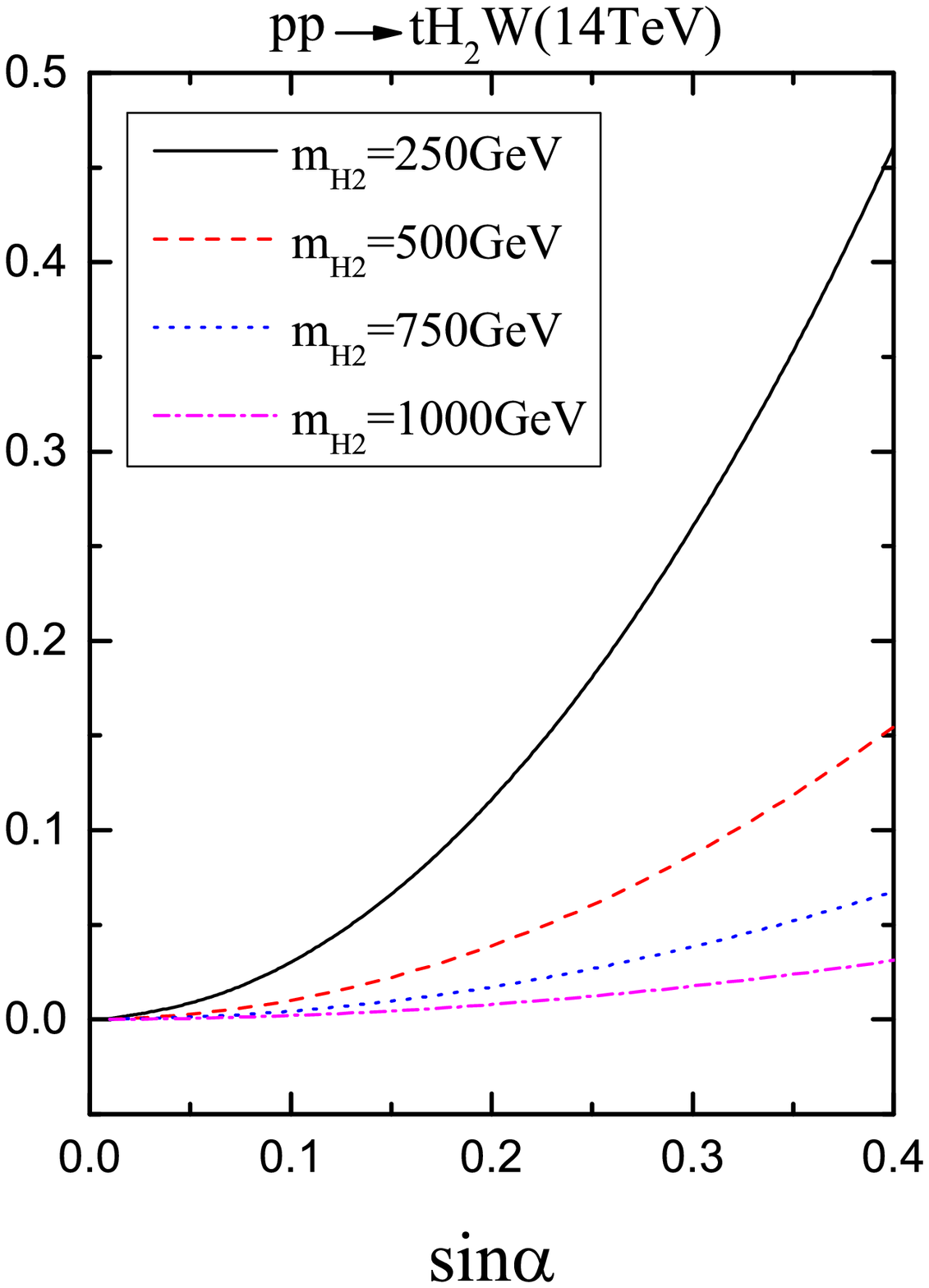}}
\vspace{-0cm}\caption{The production cross sections
$\sigma_{tH_{2}j}, \sigma_{tH_{2}b}, \sigma_{tH_{2}W}$ as a function
of $\rm sin\alpha $ at 14 TeV LHC in the $U(1)_{B-L}$
model.}\label{th2cross14}
\end{figure}

In Figs.(\ref{th2cross8}-\ref{th2cross14}), we show the production
cross sections of the processes $pp \to tH_{2}j$, $pp \to
tH_{2}\bar{b}$ and $pp \to tH_{2}W^{-}$ as a function of $\rm
sin\alpha $ at the 8 and 14 TeV LHC in the $U(1)_{B-L}$ model,
respectively. In order to see the influence of the heavy Higgs mass
$m_{H_{2}}$ on the production cross sections, we take
$m_{H_{2}}=250, 500, 750, 1000\rm GeV$ as example. We can see that
the cross sections increase with increasing $\sin\alpha$, which is
because the heavy Higgs $H_{2}$ couplings in Eq.(\ref{int}) are
proportional to $\sin\alpha$  so that the cross sections are
proportional to $\sin^{2}\alpha$.

\subsection{Observability of $pp \to tH_{2}j$}
\begin{figure}[htbp]
\scalebox{0.4}{\epsfig{file=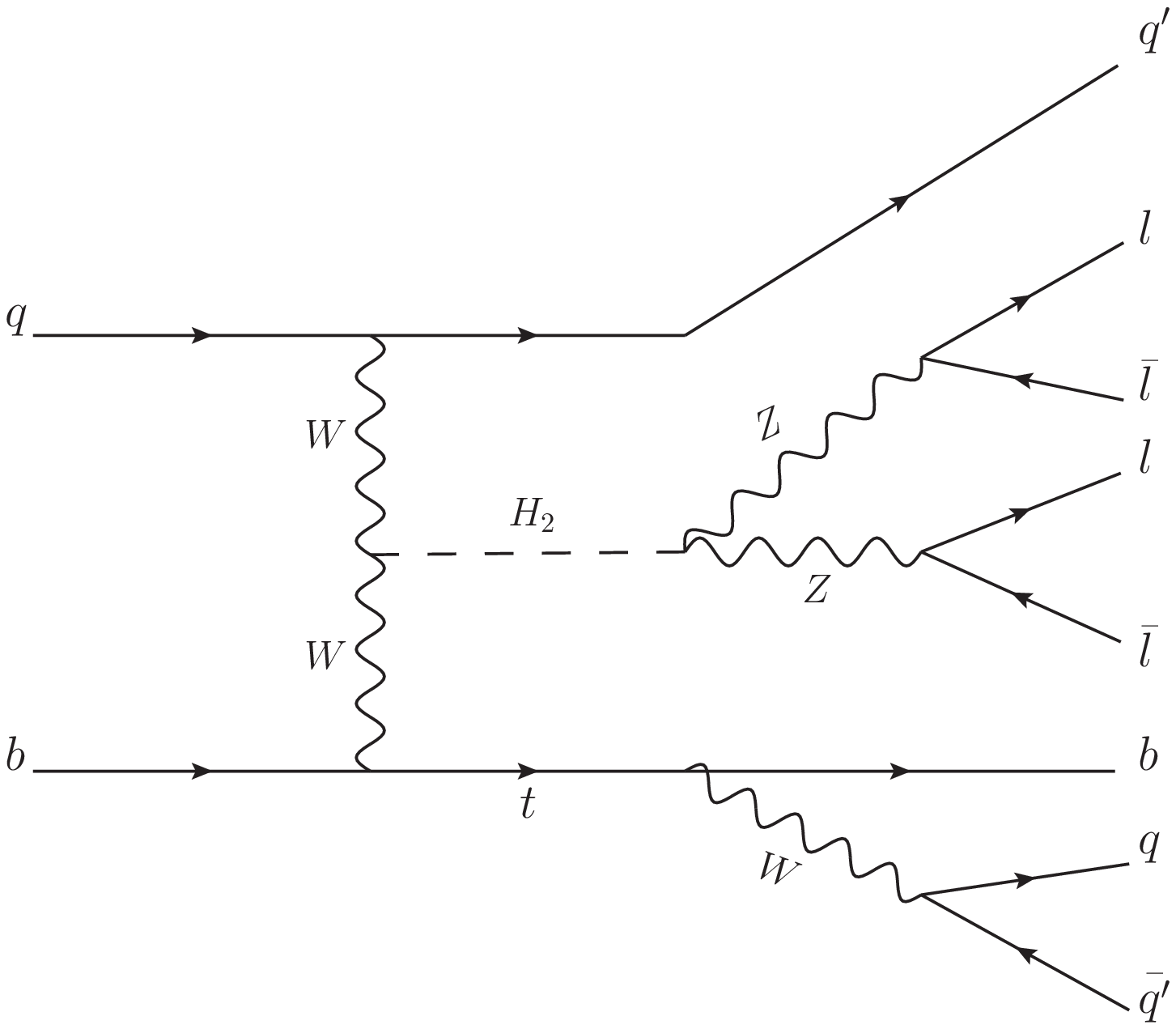}}\hspace{1cm}
\scalebox{0.4}{\epsfig{file=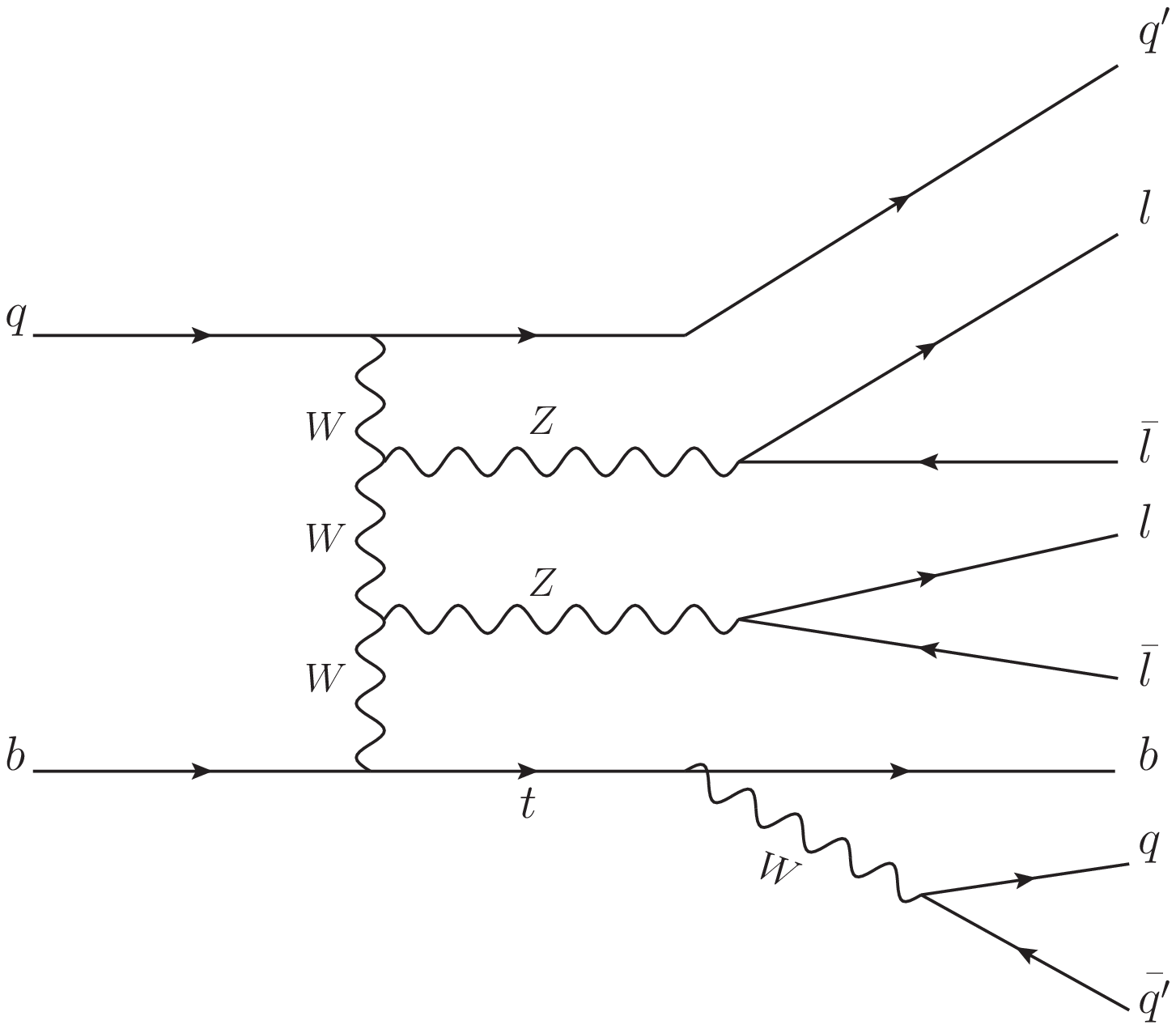}}
\vspace{-0.5cm}\caption{Feynman diagrams for signal $pp\rightarrow
tH_{2}j$ (Left) and background $pp\rightarrow tZZj$(Right) including
the decay chain with hadronic top quark, leptonic $Z$ boson decay
and Higgs decay $H_{2}\rightarrow ZZ\rightarrow
\ell_{1}^{+}\ell_{1}^{-}\ell_{2}^{+}\ell_{2}^{-}$ at the
LHC.}\label{decay}
\end{figure}

The $t$-channel process dominates amongst these three production
modes at the LHC, so we will explore the observability through the
$t$-channel $pp \to tH_{2}j$ at 14 TeV LHC in the following section.
The three most dominant decay modes of the heavy Higgs $H_{2}$ are
$WW, H_{1}H_{1}$ and $ZZ$\cite{BL-higgs}. Though the branching
fraction of $H_{2}\rightarrow ZZ$ is smaller than the branching
fractions of $H_{2}\rightarrow WW$ and $H_{2}\rightarrow
H_{1}H_{1}$, the $ZZ$ signal is much easier to separate from SM
backgrounds. For the $ZZ$ decay modes, the leptonic decay mode of
$ZZ$ offer the cleanest possible signatures though the di-jet and
semi-leptonic decay modes of $ZZ$ are larger. This leptonic decay
mode has been studied in the heavy Higgs production at the LHC and
it found that a heavy Higgs boson of mass smaller than 500 GeV can
be discovered at the LHC with high-luminosity
(HL-LHC)\cite{constraint2}. In our work, we concentrate on the
channel $pp\rightarrow t(\rightarrow W^{+}b\rightarrow q\bar{q'}
b)H_{2}(\rightarrow ZZ\rightarrow
\ell_{1}^{+}\ell_{1}^{-}\ell_{2}^{+}\ell_{2}^{-})j$ as shown in
Fig.\ref{decay}, where $H_{2}$ decays to two $Z$ bosons and the two
$Z$ bosons subsequently decay to four leptons. The signal is
characterised by
\begin{equation}
3\rm~jet + b~\rm jet + 4\ell
\end{equation}
where $j$ denotes the light jets and $\ell=e,\mu$. The largest
background for this process comes from the $tZZj$ production mode
that will generate the same final state.

\begin{figure}[htbp]
\scalebox{0.33}{\epsfig{file=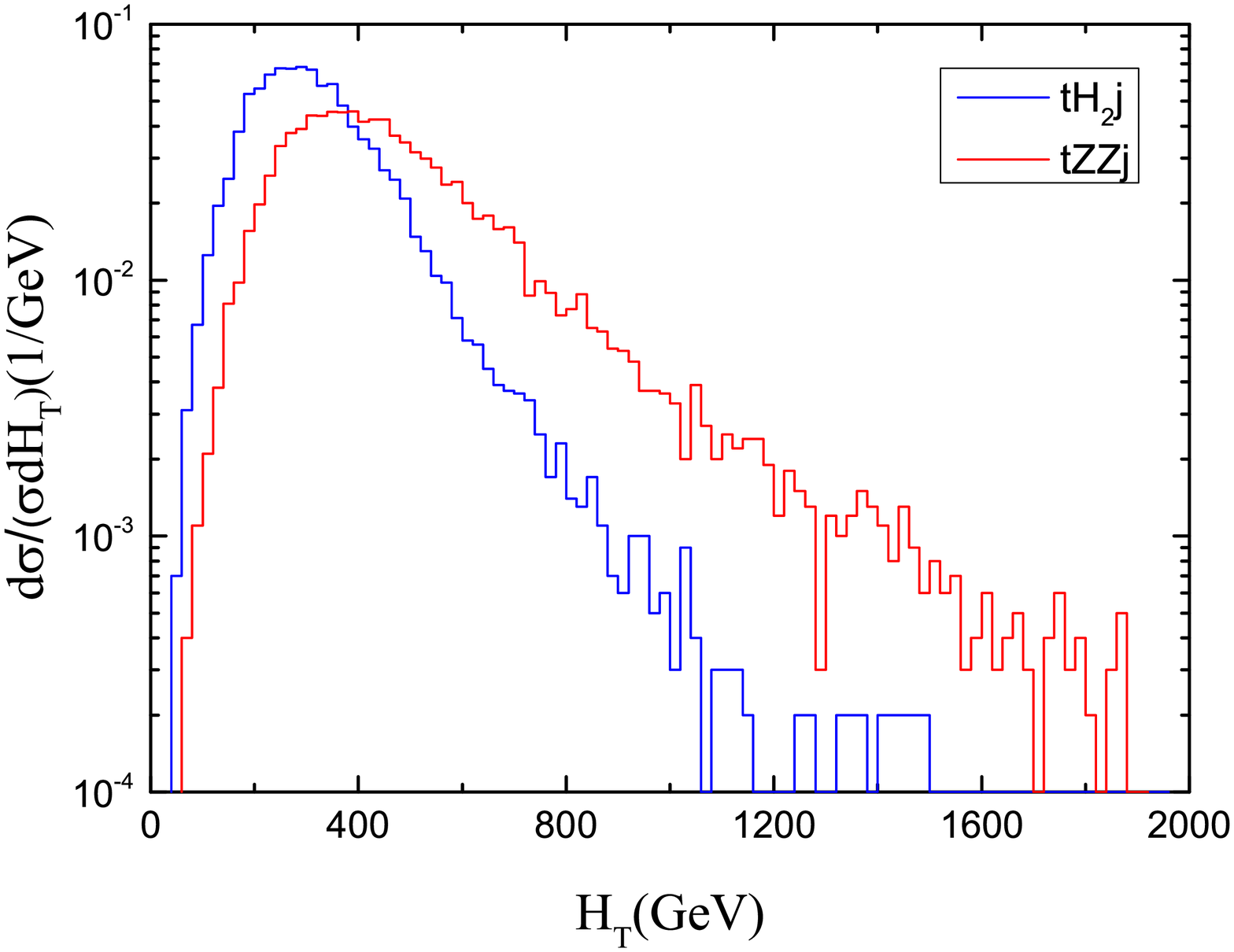}}\hspace{0cm}
\scalebox{0.33}{\epsfig{file=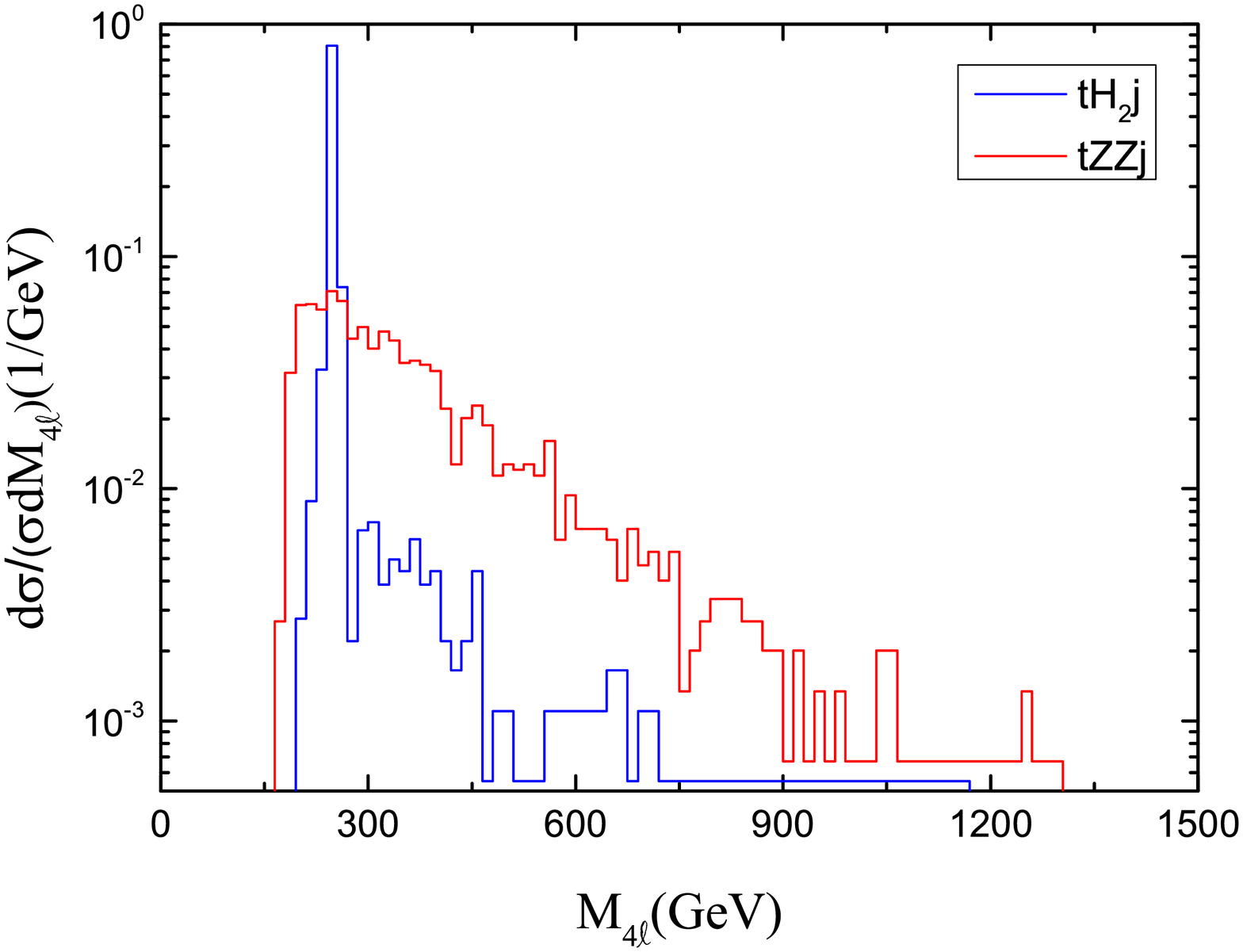}}\hspace{0.cm} \caption{The
normalized distributions of $ H_{T}, M_{4\ell}$ in the signal and
background at 14 TeV LHC for $m_{H2}=250$ GeV, $\rm
sin\alpha=0.3$.}\label{th2}
\end{figure}
We generate the signal and background events with with
\textsf{MadGraph5}\cite{MadGraph5} and perform the parton shower and
the fast detector simulations with \textsf{PYTHIA}\cite{PYTHIA} and
\textsf{Delphes}\cite{Delphes}. To simulate $b$-tagging, we take
moderate single $b$-tagging efficiency $\epsilon_{b} = 0.7$ for
$b$-jet in the final state. Follow the analysis on $t\bar{t}h$
signature by ATLAS and CMS collaborations\cite{tth-exp} at the LHC
Run-I, the events are selected to satisfy the criteria as follows:
\begin{eqnarray}\label{basic}
\nonumber\Delta R_{ik} &>&  0.4\ ,\quad  i,k = b,j \ \text{or}\  \ell \\
 p_{T}^b &>& 25 \ \text{GeV},\quad |\eta_b| <2.5 \\
\nonumber  p_{T}^\ell &>& 10 \ \text{GeV}, \quad  |\eta_\ell|<2.5  \\
\nonumber p_{T}^j &>& 25 \ \text{GeV},\quad  |\eta_j|<5.
\end{eqnarray}

\begin{table}[ht!]
\fontsize{12pt}{8pt}\selectfont

\caption{Cutflow of the cross sections for the signal and
backgrounds at 14 TeV LHC on the benchmark point ($m_{H_{2}}=250$
GeV, $\rm sin\alpha=0.3$). All the conjugate processes of the signal
and background have been included.\label{cutflow}}
\begin{center}
\newcolumntype{C}[1]{>{\centering\let\newline\\\arraybackslash\hspace{0pt}}m{#1}}
{\renewcommand{\arraystretch}{1.5}
\begin{tabular}{ C{0.3cm} C{0.3cm} C{0.cm} |C{2cm}|C{2cm}|C{2cm}| C{2cm}}
\cline{1-6} \hline
&\multicolumn{2}{c|}{\multirow{3}{*}{Cuts}}&\multicolumn{2}{c|}{$\sigma(\times
10^{-4}$fb)}
&\multicolumn{1}{c|}{\multirow{2}{*}{$\frac{S}{\sqrt{B}}$}}&\multicolumn{1}{c}{\multirow{2}{*}{$\frac{S}{B}$}}\\\cline{4-5}
&&&\multicolumn{1}{c|}{Signal} &\multicolumn{1}{c|}{Background}&&
\\\cline{4-7}
&&&\multicolumn{1}{c|}{$tH_{2}j$} &\multicolumn{1}{c|}{$tZZj$}&
3000fb$^{-1}$&
\\\hline \multicolumn{2}{c}{\multirow{1}{*}{No cuts}}&&34.3&103.9&1.84&0.33\\\cline{1-7}
\multicolumn{2}{c}{\multirow{1}{*}{Basic cuts}}&&15.6
&50.2&1.20&0.31
\\\hline \multicolumn{2}{c}{\multirow{1}{*}{$
H_{T}<
380\text{GeV}$}}&&11.3&19.5&1.40&0.58\\
\hline\multicolumn{2}{c}{\multirow{1}{*}{$|M_{4l}-250|<20
\text{GeV}$}}&&2.97&1.04&1.60&2.86\\\hline
\end{tabular}}
\end{center}
\end{table}

Due to the small signal cross section, this process has a low
signal-to-background ratio $S/\sqrt{B}$ at the LHC. In this case, we
will focus on enhancing the systematic significance $S/B$.
Considering the transverse momentum of the leptons have little
effect on the signal-to-background ratio and the systematic
significance, we don't use them as selection cuts here. After
analysis, we will adopt the following two cuts, the relevant
normalized distributions of the kinematic variables for
$m_{H_{2}}=250$ GeV, $\rm sin\alpha=0.3$ with respect to the
background are shown in Fig.\ref{th2}.

Firstly, we impose the cut $H_{T}<380$ GeV to separate signal from
background, where $H_{T}(= \sum_{\text{hadronic particles}}
\big|\big| \vec p_T \big| \big|)$ is the total transverse hadronic
energy. This cut can improve both the signal-to-background ratio
$S/\sqrt{B}$ and the systematic significance $S/B$.

After that, we apply the invariant mass of the four lepton system to
further isolate the signal and let $M_{4l}$ lie in the range
$m_{H_{2}}\pm 20$ GeV. We can see that the signal-to-background
ratio $S/\sqrt{B}$ is improved and the systematic significance $S/B$
is enhanced obviously.

The cut-flow cross sections of the signal and background for 14 TeV
LHC are summarized in Table.\ref{cutflow}. After all cuts above, we
can see that the systematic significance $S/B$ is substantially
improved. For the HL-LHC with a final integrated luminosity of
$\mathcal L=3000$fb$^{-1}$, the signal-to-background ratio
$S/\sqrt{B}$ can reach 1.6$\sigma$ and systematic significance $S/B$
can reach 2.86 for $m_{H_{2}}= 250$ GeV, $\rm sin\alpha=0.3$.
Unfortunatel, we can see that the number of signal events is very
small because of the small leptonic branching ratio of the $Z$
boson, which will be a trouble for detecting this signal at the LHC.

\section{Summary}

In the minimal $B-L$ extension of the SM, we investigated the single
top and Higgs associated production at the LHC. We computed the
production cross sections of the processes $pp\rightarrow
tH_{1}(H_{2})X(X=j,b,W)$ for 8, 14 TeV LHC and displayed the
dependance of the cross sections on the relevant $U(1)_{B-L}$ model
parameter. Moreover, we investigated the observability of process
$pp\rightarrow tH_{2}j$ followed by the decays $t\rightarrow
q\bar{q'} b$ and $H_{2}(\rightarrow ZZ\rightarrow
\ell_{1}^{+}\ell_{1}^{-}\ell_{2}^{+}\ell_{2}^{-})$ at 14 TeV LHC for
$m_{H_{2}}=250$ GeV, $\rm sin\alpha=0.3$. We performed a simple
parton-level simulation and found that it is challenging for the 14
TeV LHC and future HL-LHC with the integrated luminosity $\mathcal
L=3000$fb$^{-1}$ to observe the effect of the process $pp\rightarrow
tH_{2}j$ through this final state. So, we have to expect a collider
with higher energy and higher luminosity to probe this effect.
Maybe, a 100 TeV proton-proton collider with integrated luminosities
of 3 ab$^{-1}\sim$ 30ab$^{-1}$ can provide us a potential
opportunity\cite{100tev}.

\section*{Acknowledgement}
This work was supported by the National Natural Science Foundation
of China (NNSFC) under grants No.11405047, the Startup Foundation
for Doctors of Henan Normal University under Grant No.qd15207, the
Joint Funds of the National Natural Science Foundation of China
(U1404113), the Education Department Foundation of Henan
Province(14A140010) , the Aid Project for the Mainstay Young
Teachers in Henan Provincial Institutions of Higher Education of
China(2014GGJS-283) and Colleges and universities in Henan province
key scientific research project for 2016(16B140002).


\end{document}